\documentclass{article}%
\usepackage{amsmath}
\usepackage{amsfonts}
\usepackage{amssymb}
\usepackage{graphicx}
\usepackage{float}%
\providecommand{\U}[1]{\protect\rule{.1in}{.1in}}
\begin{document}

\begin{center}
\bigskip{\Huge Unavoidable Loop-Induced Quintessence -- Higgs Mixing and Its Phenomenological
Consequences}

{\Huge \bigskip}

{\Huge \bigskip}

\textbf{Z. Kepuladze}\footnote{zurab.kepuladze.1@iliauni.edu.ge,
zkepuladze@yahoo.com}

\bigskip

\textit{Institute of Theoretical Physics, Ilia State University, 0162 Tbilisi,
Georgia\ \vspace{0pt}\\[0pt]}

\textit{and} \textit{Andronikashvili} \textit{Institute of Physics, 0177
Tbilisi, Georgia\ }

\bigskip

\bigskip

\textbf{Abstract}
\end{center}

We investigate a class of quintessence models in which the dark-energy scalar
field interacts with sterile neutrinos responsible for neutrino mass
generation through the seesaw mechanism. Radiative corrections involving
sterile neutrinos induce Higgs--quintessence mixing and provide an effective
portal between quintessence and Standard Model particles. We calculate the
corresponding one-loop mixing amplitude and show that its structure depends on
the relation between the characteristic momentum transfer and the
sterile-neutrino mass. In the low-momentum regime the mixing is effectively
kinetic, while at high energies it acquires a mass-mixing form. A notable
result is that the overall suppression of the induced mixing is governed by
the physical neutrino mass scale, leading to a predictive relation between
neutrino properties and Higgs--quintessence interactions. The resulting
loop-induced effects are found to be suppressed by factors comparable to those
controlling the tree-level quintessence--neutrino interaction. The induced
mixing generates effective couplings of quintessence to Standard Model
fermions and gauge bosons, leading to modified Higgs, W and Z boson processes
and opening decay channels for the quintessence field, which may prove to be
cosmologically important. While the corresponding effects remain well below
current experimental sensitivities in the minimal heavy-seesaw scenario, the
framework establishes a direct connection between dark-energy dynamics,
neutrino mass generation and Higgs-sector phenomenology and provides a basis
for studying scenarios with lighter sterile neutrino states where observable
effects may be enhanced.

%%%%%%%%%%%%%%%%%%%%%%%%%%%%%%%%%%%%%%%%%%%%%%%%%%%%%%
\thispagestyle{empty}\newpage

\bigskip

\section{Introduction}

The modeling of dark energy (DE) remains one of the most intriguing yet least
understood topics at the intersection of particle physics and cosmology. While
certain aspects -- such as its energy density, evolution, and equation of
state -- are relatively well characterized, its fundamental nature and origin
remain speculative. Much like inflation, DE is often modeled
phenomenologically through the dynamics of a scalar field, commonly referred
to as quintessence \cite{PR, ZWS, BIN, BM, DS}. In some approaches, this field
is identified with the inflaton, leading to the notion of \textquotedblleft
inflessence\textquotedblright\ \cite{Gonz, Salo, Park}.\ Other frameworks
achieve the desired scalar dynamics via interactions with matter, typically
dark matter and/or neutrinos. Such interacting models have gained traction
following recent results reported by Dessi.

Models in which quintessence mimics late-time DE behavior through coupling to
neutrinos are known as mass-varying neutrino (MaVaN) scenarios \cite{FNW,
BBMT, BBBHM, MT}. MaVaN scenarios are also viable within the inflessence
framework \cite{KM, Kep, Kep1}. Linking scalar field dynamics with neutrino
physics provides insight into two persistent puzzles: the timing of DE
activation (the coincidence problem) and its small energy scale, naturally
aligned with the tiny neutrino masses. These features make the framework worth
exploring, even if it ultimately proves to be more of a theoretical exercise.
In MaVaN models, the quintessence field typically settles into a slow-roll
regime -- an adiabatic equilibrium -- where its behavior resembles DE. Yet
these constructions are highly phenomenological, differing in how DE-like
behavior is achieved and in the resulting particle masses and interactions. In
principle, the quintessence mass could range from values near the present
Hubble constant $H_{0}$ up to the grand unified theory (GUT) scale, with also
diverse neutrino couplings.

From a particle physics perspective, incorporating mechanisms such as the
seesaw \cite{Kep1}, gauging the quintessence field \cite{Kan}, and accounting
for radiative corrections can be essential \cite{FNW, Kep, Kep1, Bar}.
Theoretical subtleties also arise, for example, the role of one-loop
Coleman--Weinberg corrections. These are universally recognized as important,
yet sometimes mishandled, with undue focus on divergences that are properly
addressed within renormalizable theories. Such attempts to realisticely
describe the quintessence can generate a variety of non-standard neutrino
interactions and rare decay channels. Some of these phenomenology are
described in literature \cite{Babu, Ami, Brd}.

When the seesaw mechanism is included, Standard Model (SM) symmetries must be
respected. This requires the left-handed lepton doublet to couple to the Higgs
doublet in order to contract SU(2)$_{L}$ indices, producing a Dirac-type mass
term for neutrinos after electroweak symmetry breaking, as in Type I seesaw
frameworks. Consequently, interference between the Higgs and quintessence
fields becomes unavoidable. In this work, we aim to explore how these scalar
fields influence one another's dynamics, identify possible particle physics
and cosmological signatures, and establish bounds on such effects.

To probe the mechanism, we consider a simplified case with one heavy sterile
and one active neutrino. While this setup captures the essential nature and
dynamics of the processes, certain aspects, such as flavor-changing effects,
cannot be characterized within such a framework. Recognizing these
limitations, we nevertheless proceed with the outlined plan, as a complete
schematic description of all processes lies beyond the scope of a single
paper. Although these effects are naturally expected to be suppressed -- given
the need for the quintessence field to remain sufficiently sterile and stable
-- their exploration is nonetheless essential.

\bigskip

\section{Quintessence -- neutrino coupling structure}

Given that $y$ and $Y$ denote Yukawa couplings to the Higgs and quintessence
fields respectively, the SM Lagrangian can be extended in the following way:%
\begin{align}
L_{ext}  &  =L_{K}(N,\varphi)-V(m_{\varphi},\varphi)-\frac{y(V_{EW}+h)}%
{\sqrt{2}}\overline{N}P_{L}\nu-h.c.-\frac{(M+Y\varphi)}{2}\overline{N}N\\
&  =L_{K}(N,\varphi)-V(m_{\varphi},\varphi)-m_{D}(1+\frac{h}{V_{EW}}%
)\overline{N}P_{L}\nu-h.c.-\frac{(M+Y\varphi)}{2}\overline{N}N \label{Lext}%
\end{align}
Here, $L_{K}$ denotes the standard kinetic term of the Lagrangian,
$V(m_{\varphi},\varphi)$ is the potential of the quintessence field $\varphi$
with mass $m_{\varphi}$, $N$ represents the sterile neutrino, $m_{D}$ is the
Dirac neutrino mass, $M$ is the sterile neutrino mass, $h$ is the Higgs field,
and $V_{EW}\approx246GeV$ is the electroweak vacuum expectation value. $\nu$
is a flavor neutrino state. $P_{L,R}$ are spinor projection operators. As
discussed earlier, the range of acceptable values for $m_{\varphi}$ is broad.
The parameters $m_{D}$ and $M$ are not tightly constrained either, though they
are correlated: together they define the active neutrino mass scale via
$m_{D}^{2}/M\sim1eV$. A reasonable lower bound for the sterile neutrino mass
is about 1 Gev, corresponding $m_{D}$ values in the 1-10 keV range. The upper
bound for $M$ can be closer to the GUT scale rather than the electroweak
scale, thereby lifting $m_{D}$ and enhancing the Yukawa coupling to the Higgs
\footnote{\emph{We introduced }$m_{\varphi}$\emph{, }$M$\emph{, }$y$\emph{,
}$Y$\emph{\ as renormalizable constants, but }$L_{ext}$\emph{\ can also be
understood as a more general form of non-renormalizable cosmological
phenomenological models, expanded in series around the slow-roll
\textquotedblleft vacuum/frozen\textquotedblright\ value of the quintessence
field.}}\emph{.}

In this work, we will not attempt to diagonalize the neutrino states. Instead,
to fully capture the mechanism, we treat the states as they are and include
mixing effects when necessary. The relevant propagators and vertices are:%
\begin{align*}
D_{N}(p)  &  =\frac{i}{{\not p  }-M},\text{ \ }D_{\nu}(p)=\frac{i}{{\not p  }%
},\text{\ \ }D_{\varphi}(p)=\frac{i}{{p}_{\mu}^{2}-m_{\varphi}^{2}}\\
D(N  &  \rightarrow\nu)=-im_{D}P_{R}\text{, \ \ }D(\nu\rightarrow
N)=-im_{D}P_{L}\\
\mathit{V}_{hN\nu}  &  =-i\frac{m_{D}}{V_{EW}}P_{L}\text{, \ \ \ }%
\mathit{V}_{\varphi NN}=-iY
\end{align*}
where $D_{N}(p)$, $D_{\nu}(p)$, $D_{\varphi}(p)$ denote propagators of the
corresponding particles, while $D(N\rightarrow\nu)$ and\ $D(\nu\rightarrow N)$
are transition/oscillation factors. We denote vertex factors by $\mathit{V}%
_{X}$ and decay widths by $\Gamma_{X}$ decay rates of the $X$ to avoid
confusion between the two notions.

If we calculate oscillation effects in the propagator, the off-diagonal
propagator for a flavor neutrino oscillating into a sterile neutrino is:%
\begin{align*}
D_{\nu\rightarrow N}  &  =-\frac{i}{{\not p  }-M}im_{D}P_{L}\frac{i}{{\not p
}}+\frac{i}{{\not p  }-M}im_{D}P_{L}\frac{i}{{\not p  }}im_{D}P_{R}\frac
{i}{{\not p  }-M}im_{D}P_{L}\frac{i}{{\not p  }}+...\\
&  =\frac{im_{D}}{{\not p  }}\frac{{\not p  }+M}{{p}_{\mu}^{2}-M^{2}-m_{D}%
^{2}}P_{R}%
\end{align*}
We see that the difference between fully summing all oscillations and
considering only a single oscillation is of order $m_{D}^{2}/M^{2}$. Thus,
accounting for one oscillation captures the essence of the process with good
reliability, motivating us to restrict attention to single transitions in the
processes considered below.

\bigskip

\section{Tree processes}

The two immediate processes that come to mind are the decays of the Higgs
boson and the quintessence field into flavor neutrinos. Quintessence does not
couple directly to flavor neutrinos, but it couples to sterile neutrinos,
which can then oscillate into flavor states. The Higgs boson, by contrast, can
decay into both flavor and sterile neutrinos, with the sterile component
subsequently oscillating into flavor neutrinos. However, this Higgs decay
channel follows directly from the seesaw mechanism and is not intrinsically
tied to quintessence physics.

Nevertheless, under the assumptions of (\ref{Lext}) the matrix elements for
these processes vanish, implying that such decays do not occur. This is
straightforward to see if we denote the momentum of the Higgs/quintessence by
$k_{\mu}$ and the neutrino momenta by $p_{1\mu}$ and $p_{2\mu}$:%
\begin{align}
\mathcal{M}_{h\nu\nu}  &  \sim\overline{\nu}(p_{2})P_{R}\frac{1}{\not p
_{2}-M}P_{L}\nu(p_{1})=0\\
\mathcal{M}_{\varphi\nu\nu}  &  \sim\overline{\nu}(p_{2})P_{R}\frac{1}{\not p
_{2}-M}\frac{1}{\not p  _{1}-M}P_{L}\nu(p_{1})=0
\end{align}
with $\nu(p)$ denoting the Dirac spinor. This cancellation arises from the
chiral structure of the matrix element, together with the condition $\not p
\nu(p)=0$.

The decay rates of the Higgs boson into the flavor plus sterile neutrino, or
into two sterile neutrinos (when kinematically allowed), are not zero.
However, they are purely seesaw specific and remain unaffected by the
inclusion of quintessence. Quintessence itself can also decay into sterile
neutrinos if kinematically allowed. In most scenarios this channel is
forbidden, but when it is open the decay rate is approximately (usual formula
for boson decay into light fermion states)
\begin{equation}
\Gamma_{\varphi\rightarrow NN}\sim\dfrac{Y^{2}}{8\pi}m_{\varphi} \label{phiNN}%
\end{equation}
which enforces an extremely suppressed Yukawa coupling if quintessence
stability is paramount.

\subsection{Higgs tree-level decay into quintessence and neutrinos}

\begin{figure}[H]
\caption{The Higgs tree level non-standard decay}%
\label{fig:hdecay}
\centering
\includegraphics[width=1.1\linewidth]{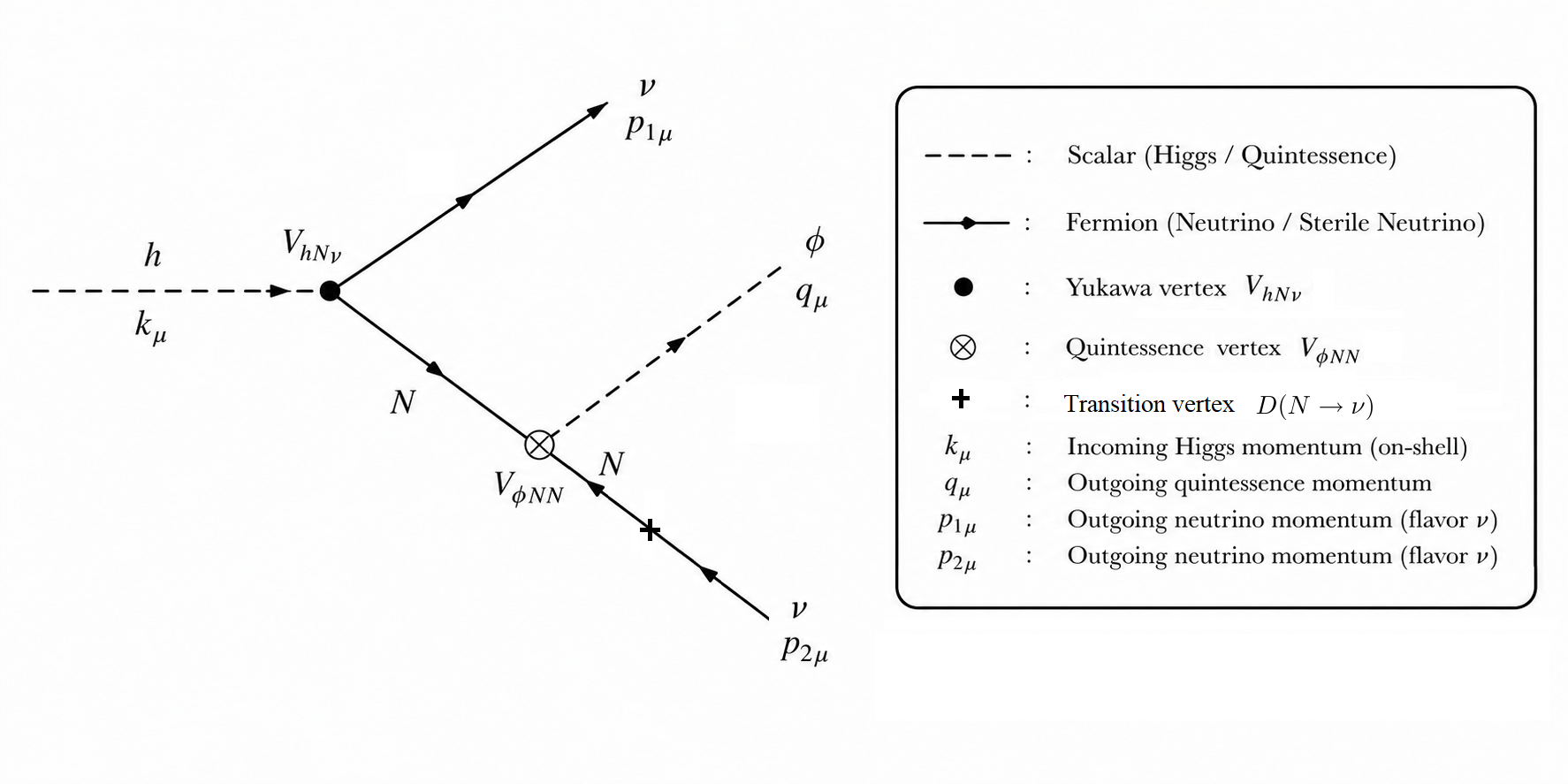}\end{figure}

At tree level, the Higgs boson can decay into quintessence and two flavor
neutrinos, provided the quintessence is light enough for the process to be
kinematically allowed. This channel is expected to be suppressed by both the
electroweak scale and the quintessence Yukawa coupling: $\mathbf{h\rightarrow
2\nu+\varphi}$. The matrix element corresponding to the diagram in
Fig.~\ref{fig:hdecay}

\noindent is the following expression
\begin{equation}
\mathcal{M}_{h\rightarrow\varphi\nu\nu}=\frac{-iYm_{D}^{2}}{V_{EW}}%
\overline{\nu}(p_{2})P_{R}\frac{1}{\not p _{2}-M}\frac{1}{\not k -\not p
_{1}-M}P_{L}\nu(p_{1})
\end{equation}
which simplifies to
\begin{equation}
\mathcal{M}_{h\rightarrow\varphi\nu\nu}=\frac{-iYm_{D}^{2}}{V_{EW}M}%
\frac{\overline{\nu}(p_{2})P_{R}\not k \nu(p_{1})}{ (k_{\mu}-p_{1\mu
})^{2}-M^{2} }\approx\frac{-iYm_{\nu}}{V_{EW}}\frac{\overline{\nu
}(p_{2})P_{R}\not k \nu(p_{1})}{  (k_{\mu}-p_{1\mu})^{2}-M^{2} }%
\end{equation}
with $(k_{\mu}-p_{1\mu})^{2}=(k_{\mu}-p_{1\mu})(k^{\mu}-p_{1}^{\mu})$ and, more
generally, we define for brevity $k_{\mu}^{2}=k_{\mu}k^{\mu}$.

Squaring the amplitude yields
\begin{equation}
\left\vert \mathcal{M}_{h\rightarrow\varphi\nu\nu}\right\vert ^{2}%
=\frac{2Y^{2}m_{\nu}^{2}}{V_{EW}^{2}}\frac{2(p_{1\alpha}k^{\alpha
})(p_{2\lambda}k^{\lambda})-M_{h}^{2}(p_{1\alpha}p_{2}^{\alpha})}{\left(
(k_{\mu}-p_{1\mu})^{2}-M^{2}\right)  ^{2}}%
\end{equation}

The differential decay rate is then defined as
\begin{equation}
d\Gamma=\frac{(2\pi)^{4}}{2k_{0}}\left\vert \mathcal{M}_{h\rightarrow
\varphi\nu\nu}\right\vert ^{2}\delta^{4}(k-q-p_{1}-p_{2})\frac{d^{3}q}%
{(2\pi)^{3}2q_{0}}\frac{d^{3}p_{1}}{(2\pi)^{3}2E_{1}}\frac{d^{3}p_{2}}%
{(2\pi)^{3}2E_{2}} \label{DR}%
\end{equation}
where $q_{\mu}=(q_{0},q_{i})$, $k_{1\mu}=(E_{1},p_{1i})$,\ $k_{2\mu}%
=(E_{2},p_{2i})$.

We can consider the Higgs decay rate in the rest frame of the Higgs boson,
since results can always be transformed to a general reference frame. In this
case,
\begin{equation}
k_{\mu}=(M_{h},0,0,0)\text{, \ \ }2p_{1\alpha}p_{2}^{\alpha}=(k_{\mu}-q_{\mu
})^{2}=M_{h}^{2}+m_{\varphi}^{2}-2M_{h}q_{0}%
\end{equation}
The squared matrix element then takes the form
\begin{equation}
\left\vert \mathcal{M}_{h\rightarrow\varphi\nu\nu}\right\vert ^{2}=\frac
{Y^{2}m_{\nu}^{2}}{V_{EW}^{2}}\frac{M_{h}^{2}(4E_{1}E_{2}-M_{h}^{2}%
-m_{\varphi}^{2}+2M_{h}q_{0})}{\left(  (k_{\mu}-p_{1\mu})^{2}-M^{2}\right)
^{2}}%
\end{equation}
Integration over the phase space in $q$, enforces momentum conservation:%
\begin{align}
d\Gamma &  =\frac{\left\vert \mathcal{M}_{h\rightarrow\varphi\nu\nu
}\right\vert ^{2}}{16M_{h}(2\pi)^{5}}\delta(k_{0}-q_{0}-E_{1}-E_{2}%
)\frac{d^{3}p_{2}d^{3}p_{1}}{q_{0}E_{1}E_{2}}\\
&  =\frac{\left\vert \mathcal{M}_{h\rightarrow\varphi\nu\nu}\right\vert ^{2}%
}{16M_{h}(2\pi)^{5}}\delta(k_{0}-q_{0}-E_{1}-E_{2})\frac{d^{3}p_{1}}{E_{1}%
}\frac{E_{2}dE_{2}d\Omega}{q_{0}}%
\end{align}
Further integration with respect to $E_{2}$ enforces energy conservation:%
\begin{align}
d\Gamma & =\frac{1}{16M_{h}(2\pi)^{5}}\frac{Y^{2}m_{\nu}^{2}M_{h}^{2}}%
{V_{EW}^{2}\left(  (k_{\mu}-p_{1\mu})^{2}-M^{2}\right)  ^{2}}\frac{d^{3}p_{1}%
}{E_{1}}\times\nonumber\\
& \frac{(4E_{1}E_{2}-M_{h}^{2}-m_{\varphi}^{2}+2M_{h}q_{0})E_{2}d\Omega}%
{M_{h}-E_{1}(1-\cos\theta)}%
\end{align}
Here, $\theta$ is the angle between $p_{1\alpha}$ and $p_{2\alpha}$. Solving
for $E_{2}$
\begin{equation}
E_{2}=\frac{M_{h}^{2}-2M_{h}E_{1}-m_{\varphi}^{2}}{M_{h}-E_{1}(1-\cos\theta)}%
\end{equation}
Its value ranges from zero up to
\begin{equation}
E_{m}=\dfrac{M_{h}^{2}-m_{\varphi}^{2}}{2M_{h}}\text{, }\theta\in\lbrack0,\pi]
\end{equation}
With this, the phase space integration over $\theta$ yields
\begin{equation}
\Gamma=\frac{Y^{2}m_{\nu}^{2}}{8(2\pi)^{3}V_{EW}^{2}}\int_{0}^{E_{m}}%
\frac{\left(  M_{h}^{2}-2M_{h}E_{1}+m_{\varphi}^{2}\right)  ^{2}}{\left(
M_{h}^{2}-2M_{h}E_{1}\right)  \left(  M_{h}^{2}-2M_{h}E_{1}-M^{2}\right)
^{2}}E_{1}^{2}dE_{1}%
\end{equation}

For simplicity, let us neglect the quintessence mass $m_{\varphi}$. When
$m_{\varphi}$ is small, it is negligable and when it is not, unless
$m_{\varphi}$ approaches $M_{h}$, the effect is not order changing. Setting
$m_{\varphi}=0$ gives%

\begin{equation}
\Gamma=\frac{Y^{2}m_{\nu}^{2}}{8(2\pi)^{3}V_{EW}^{2}}\int_{0}^{E_{m}}%
\frac{\left(  M_{h}^{2}-2M_{h}E_{1}\right)  }{\left(  M_{h}^{2}-2M_{h}%
E_{1}-M^{2}\right)  ^{2}}E_{1}^{2}dE_{1}%
\end{equation}
Before integrating over $E_{1}$, note that resonance occurs when $m_{\varphi
}^{2}<M^{2}<M_{h}^{2}$, but in the heavy sterile neutrino limit $M^{2}\gg
M_{h}^{2}$, the decay rate can be approximated as
\begin{equation}
\Gamma\approx\frac{Y^{2}m_{\nu}^{2}}{128\ast48\pi^{3}V_{EW}^{2}}\frac
{M_{h}^{5}}{M^{4}}%
\end{equation}

This is extremely suppressed. For example, with $m_{\nu}\sim1$ eV and $M\sim1$ TeV%

\begin{equation}
\Gamma\sim Y^{2}10^{-21}\left(  \frac{m_{\nu}}{1\text{ eV}}\right)
^{2}\left(  \frac{1\text{ TeV}}{M}\right)  ^{4}\text{eV}%
\end{equation}
If the sterile neutrino is only slightly heavier than the Higgs boson, the
leading order rate is
\begin{equation}
\Gamma\approx\frac{Y^{2}m_{\nu}^{2}}{128\ast8\pi^{3}V_{EW}^{2}}M_{h}\sim
Y^{2}10^{-17}\text{eV}%
\end{equation}
still negligible for Higgs phenomenology.

In the resonance case, the sterile neutrino is unstable, and the propagator
must be modified:
\[
\frac{1}{(k_{\mu}-p_{1\mu})^{2}-M^{2}}\rightarrow\frac{1}{(k_{\mu}-p_{1\mu
})^{2}-M^{2}+iM\Gamma_{M}}
\]
where $\Gamma_{M}$ is the sterile neutrino decay width. The differential rate
becomes
\begin{equation}
\frac{d\Gamma}{dE_{1}}=\frac{Y^{2}m_{\nu}^{2}}{8(2\pi)^{3}V_{EW}^{2}}%
\frac{\left(  M_{h}^{2}-2M_{h}E_{1}\right)  }{\left(  M_{h}^{2}-2M_{h}%
E_{1}-M^{2}\right)  ^{2}+M^{2}\Gamma_{M}^{2}}E_{1}^{2}%
\end{equation}
At the resonance energy%
\begin{equation}
M_{h}^{2}-2M_{h}E_{1}^{res}-M^{2}=0\text{ \ }\rightarrow\text{ \ }E_{1}%
^{res}=\dfrac{M_{h}^{2}-M^{2}}{2M_{h}}%
\end{equation}
we have
\begin{equation}
\frac{d\Gamma_{res}}{dE_{1}}=\frac{Y^{2}m_{\nu}^{2}}{8(2\pi)^{3}V_{EW}^{2}%
}\frac{(M_{h}^{2}-M^{2})^{2}}{4M_{h}^{2}\Gamma_{M}^{2}}%
\end{equation}
For sterile neutrino masses around the GeV scale, this can be amplified by the
factor $M_{h}^{2}/\Gamma_{M}^{2}$, giving%
\begin{equation}
\frac{d\Gamma_{res}}{dE_{1}}\sim10^{-3}\frac{Y^{2}m_{\nu}^{2}}{\Gamma_{M}^{2}}%
\end{equation}

Integrating over the full phase space, however, shows that the total decay
rate is not significantly enhanced -- at most by an order of magnitude:
\begin{equation}
\Gamma\sim\frac{Y^{2}m_{\nu}^{2}}{64\pi^{3}V_{EW}^{2}}M_{h}%
\end{equation}
This is explained by the fact that although the resonance peak is amplified,
its narrow width prevents it from substantially affecting the integrated rate.

We can conclude that this process has no parameter space with significant
impact on Higgs physics unless the Yukawa coupling $Y$ is taken to be
uncomfortably large. In most scenarios, the effective Yukawa coupling $Y$ is
assumed to be of order unity or smaller.

\section{One loop effects}

\begin{figure}[H]
\caption{Self energy loop}%
\label{fig:loop}
\centering
\includegraphics[width=1.1\linewidth]{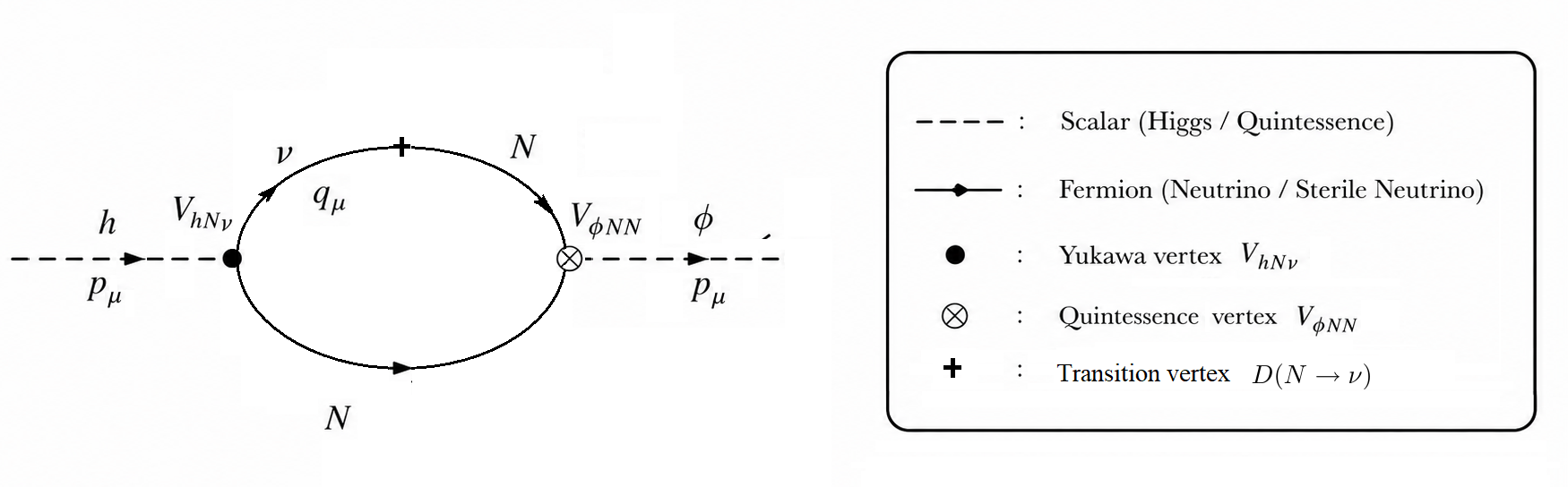}\end{figure}

The immediate physical effect that comes to mind is the mixing between the
Higgs and quintessence fields induced by neutrino loops Fig.~\ref{fig:loop} .
This polarization
loop (self-energy), which mixes the scalar fields, has the following form:%
\begin{align*}
\Pi_{h\varphi}  &  =\int\frac{d^{4}q}{(2\pi)^{4}}Tr[\frac{1}{\not q  }%
\frac{m_{D}}{V_{EW}}P_{R}\frac{1}{\not p  -\not q  -M}Y\frac{1}{\not q
-M}m_{D}P_{L}]\\
&  =2Y\frac{m_{\nu}M^{2}}{V_{EW}}\int\frac{d^{4}q}{(2\pi)^{4}}\frac{q_{\mu
}p^{\mu}}{q_{\alpha}^{2}(q_{\sigma}^{2}-M^{2})((p-q)_{\lambda}^{2}-M^{2}%
)}=2Y\frac{m_{\nu}M^{2}}{V_{EW}}p^{\mu}I_{\mu}%
\end{align*}

Using Feynman parametrization, the integral can be rewritten as%
\begin{equation}
I_{\mu}=\int_{0}^{1}dx\int_{0}^{1-x}dy\int\frac{d^{4}q}{(2\pi)^{4}}%
\frac{xp_{\mu}}{(q_{\alpha}^{2}+x(1-x)p_{\alpha}^{2}-(x+y)M^{2})^{3}}%
\end{equation}
This expression is finite. Integrating over momentum gives%
\begin{align}
I_{\mu}  &  =p_{\mu}\int_{0}^{1}xdx\int_{0}^{1-x}dy\frac{-i}{32\pi^{2}}%
\frac{1}{x(1-x)p_{\alpha}^{2}-(x+y)M^{2}}\nonumber\\
&  =\frac{ip_{\mu}}{32\pi^{2}M^{2}}\left[  \int_{0}^{1}xdx\ln[\frac
{x(1-x)p_{\alpha}^{2}-M^{2}}{(1-x)p_{\alpha}^{2}-M^{2}}]-\frac{1}{4}\right]
\end{align}
The final part of the integral can be evaluated, but the expression is not
simple. Instead, we can estimate useful limits to obtain asymptotic behavior:
\begin{align}
\text{\textbf{1. \ }}p_{\alpha}^{2}  &  \ll M^{2}\text{ }\rightarrow\text{
}I_{\mu}=\frac{ip_{\mu}}{128\pi^{2}M^{2}}\text{\ }\\
\text{\textbf{2. \ }}p_{\alpha}^{2}  &  \gg M^{2}\text{ }\rightarrow\text{
}I_{\mu}=\frac{ip_{\mu}}{32\pi^{2}p_{\alpha}^{2}}%
\end{align}
Correspondingly, the polarization function becomes
\begin{align}
\text{\textbf{\ }}p_{\alpha}^{2}  &  \ll M^{2}\text{ }\rightarrow\Pi
_{h\varphi}=i\delta p_{\alpha}^{2}\label{KM}\\
\text{\textbf{\ }}p_{\alpha}^{2}  &  \gg M^{2}\text{ }\rightarrow\Pi
_{h\varphi}=4i\delta M^{2} \label{MM}%
\end{align}
with%
\begin{equation}
\delta=\frac{1}{64\pi^{2}}\frac{Ym_{\nu}}{V_{EW}} \label{delta}%
\end{equation}

$\Pi_{h\varphi}$ has the same structure whether the Higgs oscillates into
quintessence or vice versa, when expressed in terms of $p_{\alpha}^{2}$ and
$M$ with out apprixmations. From the form of $\Pi_{h\varphi}$, we conclude
that depending on the process one can encounter either kinetic mixing, mass
mixing, or a mixed case. For on shell processes, $p_{\alpha}^{2}$ is naturally
associated with the corresponding mass scales $M_{h}$ or $m_{\varphi}$, while
for off shell effects, a wide variety of possible $p_{\alpha}^{2}$ values can
arise. Thus, we can distinguish three scenarios:

\textbf{1. }Heavy sterile neutrino case: $M\gg M_{h}$ and $M\gg m_{\varphi}$.
At TeV scales and below, regardless of the hierarchy between quintessence and
Higgs masses, the mixing is of kinetic type.

\textbf{2.} Light sterile neutrino case: $M\ll M_{h}$ and $M\ll m_{\varphi}$.
This corresponds to a clear example of mass type mixing, but is only
realizable for very heavy quintessence scenarios.

\textbf{3. }Intermediate or near-degenerate case: When the mass scales are
close to one another, or the sterile neutrino is lighter than the Higgs but
heavier than quintessence, or when the propagating momentum satisfies
$p_{\alpha}^{2}\sim M^{2}$, the full expression must be retained and mixing
effects evaluated case-by-case at the Feynman diagram level.

Let us note that the suppression structure $Ym_{\nu}/V_{EW}$ appearing in this
loop induced mixing effect is the same as in the Higgs tree level decay
process considered above.

\subsection{The kinetic mixing}

For the case\ $p_{\alpha}^{2}\ll M^{2}$, kinetic mixing remains uniform at and
below the electroweak scale. In this regime, it is straightforward to
incorporate the corresponding mixing operator into the Lagrangian:
\begin{equation}
L_{mix}=\delta\left(  \partial_{\mu}h\right)  \left(  \partial^{\mu}%
\varphi\right)
\end{equation}
Diagonalization can then be attempted by identifying mass states. The linear
transformation
\begin{equation}
h\rightarrow h+\delta\frac{m_{\varphi}^{2}}{M_{h}^{2}-m_{\varphi}^{2}}%
\varphi;\text{ \ }\varphi\rightarrow\varphi-\delta\frac{M_{h}^{2}}{M_{h}%
^{2}-m_{\varphi}^{2}}h\text{\ } \label{km1}%
\end{equation}
diagonalizes the scalar states. This procedure is equivalent to performing an
orthogonal transformation between the Higgs and quintessence scalars,
normalizing the kinetic structure, and then diagonalizing the mass matrix by
another orthogonal transformation. However, the transformation is given only
to linear order in $\delta$. In this limit, the mass values remain unchanged.
Calculating effects at order $\delta^{2}$ would not be consistent, since two
loop contributions are expected to be of the same order.

Physically, this mixing corresponds to on-shell quintessence or Higgs bosons
oscillating into their counterpart and then decaying into whatever channels
are kinematically allowed. Such mixing allows quintessence to interact with SM
particles -- lepton--antilepton pairs, quark--antiquark pairs, and massive
gauge bosons similar to the Higgs itself, but with reduced strength defined by
the transformation above. Some of this three particle interaction verteces
are:
\begin{align}
\mathit{V}_{\varphi ll,\varphi qq}  &  =i\mathrm{y}_{l,q}\delta\frac
{m_{\varphi}^{2}}{M_{h}^{2}-m_{\varphi}^{2}}\label{phill}\\
\mathit{V}_{\varphi ZZ}  &  =2i\frac{M_{Z}^{2}}{V_{EW}}\delta\frac{m_{\varphi
}^{2}}{M_{h}^{2}-m_{\varphi}^{2}};\text{ \ \ }\mathit{V}_{\varphi WW}%
=2i\frac{M_{W}^{2}}{V_{EW}}\delta\frac{m_{\varphi}^{2}}{M_{h}^{2}-m_{\varphi
}^{2}}\text{\ }\label{ZW}\\
\mathit{V}_{\varphi hh}  &  =-3i\frac{M_{h}^{2}}{V_{EW}}\delta\frac
{m_{\varphi}^{2}}{M_{h}^{2}-m_{\varphi}^{2}};\text{ \ \ } \label{hhphi}%
\end{align}
Here, $\mathrm{y}_{l,q}$ are the Yukawa couplings of leptons and quarks to the
Higgs, while $M_{Z}$ and $M_{W}$ are masses of $Z$ and charged $W$-bosons.
Although these couplings are strongly suppressed and none of them open direct
kinematic decay channels for a light quintessence field, the analogy with the
Higgs suggests that radiative decay into two photons remains the only
guaranteed possibility. This process, together with other potential
quintessence decay channels, may be relevant for cosmological signatures: a
quintessence condensate could undergo perturbative decay, emit photons of
characteristic energy, and thereby alter the dynamics of dark energy.

Finally, if the masses of quintessence and the Higgs are correlated by some
yet unknown physics, and a certain level of fine-tuning occurs between them,
the mixing effect could be significantly amplified, increasing the likelihood
of detection.

\subsubsection{$Z\rightarrow\varphi+l+\overline{l}$}

Fig.~\ref{fig:Zdecay} below illustrates how this process proceeds.
\begin{figure}[H]
\caption{Z boson invisible decay}%
\label{fig:Zdecay}
\centering
\includegraphics[width=1.1\linewidth]{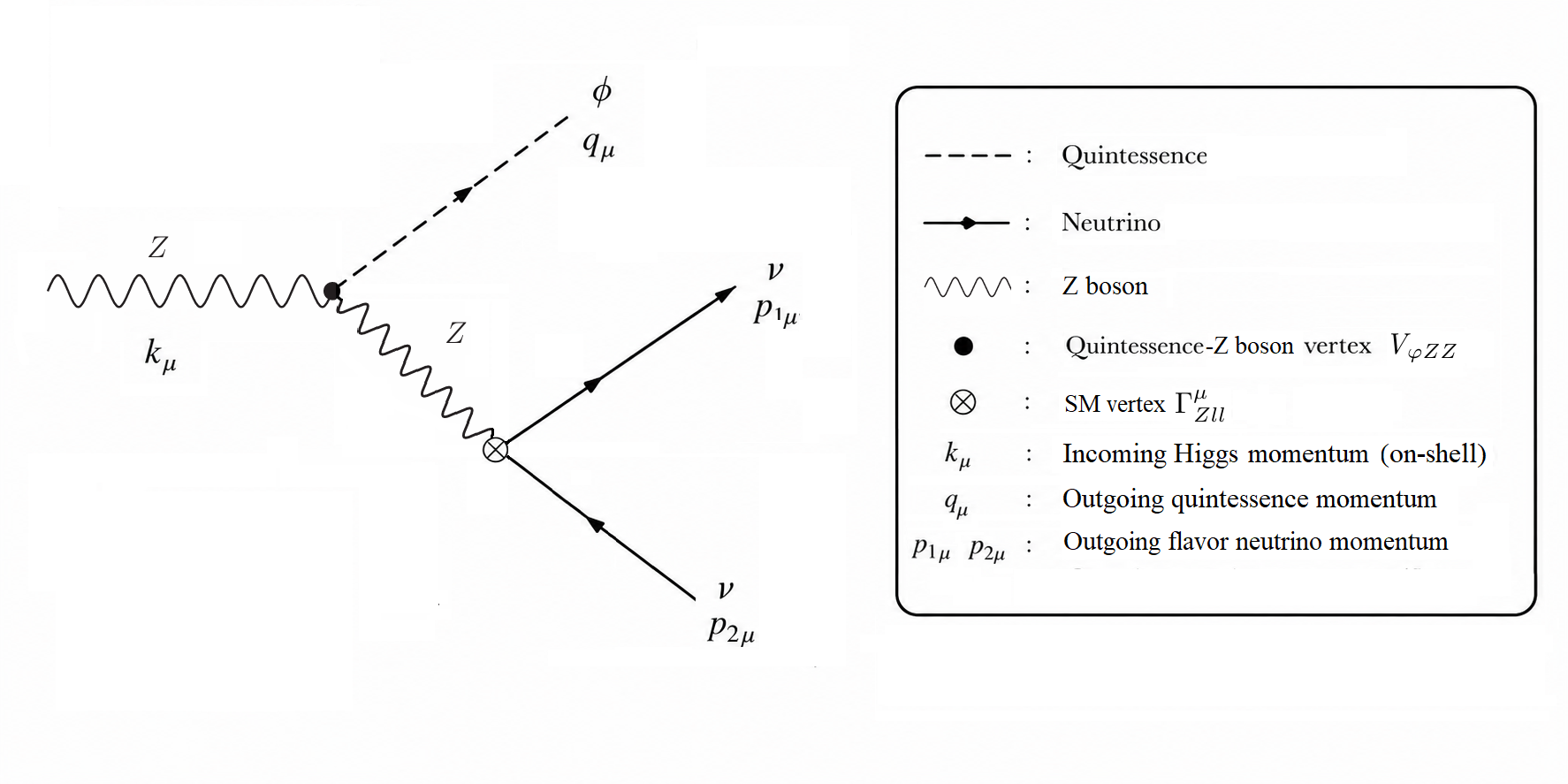}\end{figure}To calculate its
matrix element, we first recall the form of the $Z$-lepton-antilepton vertex:
\begin{equation}
\Gamma_{Zll}^{\mu}=-\frac{ie}{2\sin2\theta_{w}}\gamma^{\mu}(g_{l}-\gamma_{5})
\end{equation}
where $e$ is the electric charge, $\theta_{w}$ is Weinberg angle and $g_{l}=1$
for neutrinoes, while $g_{l}$ $=1-4\sin^{2}\theta_{w}$ for electrons, muons,
and tau leptons. In the leading order, lepton masses can be neglected.

Employing (\ref{ZW}), the matrix element becomes%
\begin{equation}
\mathcal{M}_{z\rightarrow\varphi ll}=-\frac{ie}{2\sin2\theta_{w}}\overline
{u}_{2}\gamma_{\mu}(g_{l}-\gamma_{5})v_{1}D^{\mu\nu}(k-q)2i\frac{M_{Z}^{2}%
}{V_{EW}}\delta\frac{m_{\varphi}^{2}}{M_{h}^{2}-m_{\varphi}^{2}}\xi_{\nu}%
\end{equation}
where $D^{\mu\nu}(k-q)$ is the Z-boson propagator in unitary gauge carrying
momentum $k-q$. Using on-shell conditions, this simplifies to%
\begin{equation}
\mathcal{M}_{z\rightarrow\varphi ll}=-\frac{i\delta e}{\sin2\theta_{w}}%
\frac{M_{Z}^{2}}{V_{EW}}\frac{m_{\varphi}^{2}}{M_{h}^{2}-m_{\varphi}^{2}}%
\frac{\overline{u}_{2}\not \xi (g_{l}-\gamma_{5})v_{1}}{m_{\varphi}%
^{2}-2k^{\mu}q_{\mu}}%
\end{equation}

Squaring the amplitude and averaging over Z-boson polarizations yields
\begin{align}
\left\vert \mathcal{M}_{z\rightarrow\varphi ll}\right\vert ^{2}  &
=\frac{\delta^{2}e^{2}(1+g_{l}^{2})}{3\sin^{2}2\theta_{w}}\frac{M_{Z}%
^{4}m_{\varphi}^{4}}{V_{EW}^{2}\left(  M_{h}^{2}-m_{\varphi}^{2}\right)  ^{2}%
}\times\nonumber\\
& \frac{Tr[\not p_{2}\not k\not p_{1}\not k]/M_{Z}^{2}-Tr[\not p_{2}%
\gamma^{\alpha}\not p_{1}\gamma_{\alpha}]}{\left(  m_{\varphi}^{2}-2k^{\mu
}q_{\mu}\right)  ^{2}}%
\end{align}
The Dirac traces evaluate to
\begin{align*}
Tr[\not p_{2}\not k\not p_{1}\not k]/M_{Z}^{2} &  =2\left(  k^{\mu}p_{1\mu
}\right)  \left(  k^{\mu}p_{2\mu}\right)  /M_{Z}^{2}-\left(  p_{1}^{\mu
}p_{2\mu}\right)  \\
Tr[\not p_{2}\gamma^{\alpha}\not p_{1}\gamma_{\alpha}] &  =2\left(  p_{1}%
^{\mu}p_{2\mu}\right)  -4\left(  p_{1}^{\mu}p_{2\mu}\right)
\end{align*}
Thus,
\begin{align}
\left\vert \mathcal{M}_{z\rightarrow\varphi ll}\right\vert ^{2}  &
=\frac{\delta^{2}e^{2}(1+g_{l}^{2})}{3\sin^{2}2\theta_{w}}\frac{M_{Z}%
^{4}m_{\varphi}^{4}}{V_{EW}^{2}\left(  M_{h}^{2}-m_{\varphi}^{2}\right)  ^{2}%
}\times\nonumber\\
& \frac{\left(  p_{1}^{\mu}p_{2\mu}\right)  +2\left(  k^{\mu}p_{1\mu}\right)
\left(  k^{\mu}p_{2\mu}\right)  /M_{Z}^{2}}{\left(  m_{\varphi}^{2}-2k^{\mu
}q_{\mu}\right)  ^{2}}%
\end{align}

Using momentum conservation and taking the Z-boson in its rest frame
simplifies the expression to
\begin{align}
\left\vert \mathcal{M}_{z\rightarrow\varphi ll}\right\vert ^{2}  &
=\frac{\delta^{2}e^{2}(1+g_{l}^{2})}{6\sin^{2}2\theta_{w}}\frac{M_{Z}%
^{4}m_{\varphi}^{4}}{V_{EW}^{2}\left(  M_{h}^{2}-m_{\varphi}^{2}\right)  ^{2}%
}\times\nonumber\\
& \frac{M_{Z}^{2}+m_{\varphi}^{2}-2M_{Z}q_{0}+4E_{1}E_{2}}{\left(  m_{\varphi
}^{2}-2M_{Z}q_{0}\right)  ^{2}}%
\end{align}
The differential decay rate is then
\begin{equation}
d\Gamma_{Z}^{new}=\frac{\left\vert \mathcal{M}_{z\rightarrow\varphi
ll}\right\vert ^{2}}{8M_{Z}(2\pi)^{5}}\delta^{4}(k-q-p_{1}-p_{2})\frac{d^{3}%
q}{q_{0}}\frac{d^{3}p_{1}}{E_{1}}\frac{d^{3}p_{2}}{E_{2}}%
\end{equation}
Integration over phase space proceeds analogously to the Higgs decay case
discussed earlier. Here, however, resonance does not occur. As before, we can
neglect the quintessence mass, since small $m_{\varphi}$ maximizes the weak
boson decay rate (excluding effect because of mixing ratio $\dfrac{m_{\varphi
}^{2}}{M_{h}^{2}-m_{\varphi}^{2}}$ (\ref{km1}), as it mentioned above
fine-tuned values of $M_{h}$ and $m_{\varphi}$ may leave perturabative limit
for the mixing). In the limit of small $m_{\varphi}$, in the leading order we
obtain
\begin{equation}
\Gamma_{Z}^{new}\approx\frac{\delta^{2}e^{2}(1+g_{l}^{2})}{64\pi^{3}\sin
^{2}2\theta_{w}}\frac{M_{Z}^{3}m_{\varphi}^{4}}{V_{EW}^{2}\left(  M_{h}%
^{2}-m_{\varphi}^{2}\right)  ^{2}}(\ln\frac{M_{Z}}{m_{\varphi}}-47/24+\frac
{5\pi m_{\varphi}}{24M_{Z}})
\end{equation}
Relative to the Standard Model Z-boson decay rate into a specific lepton
channel, $\Gamma_{Zll}$, the ratio is%
\begin{equation}
\frac{\Gamma_{Z}^{new}}{\Gamma_{Zll}}\sim\delta^{2}\frac{m_{\varphi}^{4}%
}{\left(  M_{h}^{2}-m_{\varphi}^{2}\right)  ^{2}}%
\end{equation}

This ratio is proportional to the square of the mixing parameter (\ref{km1}).

A similar result holds for analogous three-particle decay processes of the
charged weak boson and the Higgs boson (madiated by the (\ref{hhphi})). In all
cases, the suppression factor is too severe to have any meaningful impact on
accelerator or cosmic ray physics.

\subsubsection{The Higgs decay via quintessence self interaction}

Apart from decay processes fostered by Standard Model interactions, depending
on the model one may also encounter three point self-interactions of the
quintessence field. Through mixing, such interactions generate the possibility
of Higgs decay into two quintessence particles in light quintessence
scenarios:%
\begin{equation}
L_{int}=\mathrm{M}_{int}\varphi^{3}\text{ }\rightarrow\text{ }\mathrm{V}%
_{h\varphi\varphi}=-6i\delta\frac{M_{h}^{2}}{M_{h}^{2}-m_{\varphi}^{2}%
}\mathrm{M}_{int}%
\end{equation}
This scalar process is straightforward. In the Higgs rest frame, the decay
rate into the quintessence channel can be written as
\begin{align}
\Gamma_{h\rightarrow\varphi\varphi}  &  =\frac{1}{2}\frac{\left\vert
\mathrm{V}_{h\varphi\varphi}\right\vert ^{2}}{32\pi^{2}M_{h}}\int\delta
^{4}(k-q-p)\frac{d^{3}q}{q_{0}}\frac{d^{3}p}{p_{0}}\\
&  \approx\frac{\left\vert \mathrm{V}_{h\varphi\varphi}\right\vert ^{2}}{16\pi
M_{h}}\sqrt{1-\frac{4m_{\varphi}^{2}}{M_{h}^{2}}}\sim\frac{\delta
^{2}\mathrm{M}_{int}^{2}}{M_{h}}%
\end{align}
where $p$ and $q$ are the outgoing quintessence momenta and $k=(M_{h},0,0,0)$.
Taking (\ref{delta}) into account, and assuming an $m_{\nu}\approx1$ eV, we
obtain%
\begin{equation}
\Gamma_{h\rightarrow\varphi\varphi}\sim10^{-40}Y^{2}\mathrm{M}_{int}^{2}%
\end{equation}
This necessarily pushes $\mathrm{M}_{int}$ to scales higher than $10^{20}$ eV
in order for the decay rate to remain at least in the eV range. Such a
requirement corresponds to an extremely strong coupling. For reference, in
inflessence models where mass and coupling are defined as%
\begin{equation}
m_{\varphi}\sim\frac{V_{0}^{2}}{M_{P}}\text{; \ \ }\mathrm{M}_{int}\sim
\frac{V_{0}^{4}}{M_{P}^{3}}%
\end{equation}
with $M_{P}$ the Planck mass and $V_{0}$ the inflation scale, such a large
value of $\mathrm{M}_{int}$ pushes the inflation scale up to the GUT regime
and the inflessence mass also near the GUT scale. Consequently, this decay
becomes impossible in inflessence scenarios where the coupling is strong
enough. Beyond inflessence models, quintessence potentials and self-couplings
are typically feeble.

\subsubsection{Quintessence decay}

From the decay processes of massive bosons discussed earlier, it is clear that
the tiny contributions to SM particle decay rates play no practical role.
However, the situation for the quintessence field is different. Quintessence
is supposed to be sufficiently stable to fulfill the role of dark energy, and
even very small decay rates can have significant consequences for its
cosmological evolution. If we denote the quintessence decay rate by
$\Gamma_{\varphi}$, then the evolution of the field energy density in coupled
models can be modeled by
\begin{equation}
\dot{\rho}_{\varphi}=-(3H+\Gamma_{\varphi})(1+\omega_{\varphi})\rho_{\varphi}%
\end{equation}
where $H$ is the Hubble constant and $\omega_{\varphi}$ is the equation of
state parameter. In the slow-roll regime, even if $\omega_{\varphi}$ remains
close to $-1$ and if $H_{0}\ll\Gamma_{\varphi}$ (with $H_{0}$ the current
Hubble constant), the energy density decays exponentially:
\begin{equation}
\rho_{\varphi}\sim\exp[-(1+\omega_{\varphi})\Gamma_{\varphi}]
\end{equation}
Depending on how close $\omega_{\varphi}$ is to minus one and how large
$\Gamma_{\varphi}$ is, quintessence decay can be unacceptably fast for DE
modeling. In any case, $H_{0}$ provides a natural reference scale for
constraining $\Gamma_{\varphi}$. Models in which quintessence does not sit at
the absolute minimum of its potential, and achieves slow roll differently,
cannot afford $\Gamma_{\varphi}\gg H_{0}$. In the recent work \cite{chanda} it
is even claimed that $\Gamma_{\varphi}$ few order of magnitude smaller then
$H_{0}$ is to explain DESI results and phantom crossing. So, may be smaller
$\Gamma_{\varphi}$ is even preferable.

$\mathbf{\varphi\rightarrow e}^{-}\mathbf{+e}^{+}$. One loop mixing effects
open the possibility for quintessence decay into an electron--positron pair if
$m_{\varphi}>4m_{e}$. With mixing already taken into account, this is a simple
tree-level process with coupling strength $\delta_{e}=\delta\dfrac
{\mathrm{y}_{e}m_{\varphi}^{2}}{M_{h}^{2}-m_{\varphi}^{2}}$ (\ref{phill}). The
squared matrix element is%
\begin{equation}
\left\vert \mathcal{M}_{\varphi}\right\vert ^{2}=\delta_{e}^{2}Tr[\not q
\not p  -m_{e}^{2}]=4\delta_{e}^{2}(q_{\mu}p^{\mu}-m_{e}^{2})
\end{equation}
where $q_{\mu}$ and $p_{\mu}$ are the electron and positron four momenta.
Momentum conservation gives $q_{\mu}p^{\mu}=$ $m_{\varphi}^{2}/2-m_{e}^{2} $,
so
\begin{equation}
\left\vert \mathcal{M}_{\varphi}\right\vert ^{2}=2\delta_{e}^{2}(m_{\varphi
}^{2}-m_{e}^{2}/4)
\end{equation}

In the rest frame, the decay rate is%

\begin{align}
\Gamma_{\varphi\rightarrow ee}  &  =\frac{\delta_{e}^{2}(m_{\varphi}^{2}%
-m_{e}^{2}/4)}{16\pi^{2}m_{\varphi}}\int\delta^{4}(k-q-p)\frac{d^{3}q}{q_{0}%
}\frac{d^{3}p}{p_{0}}\\
&  =\frac{\delta_{e}^{2}(m_{\varphi}^{2}-m_{e}^{2}/4)}{4\pi m_{\varphi}}%
\sqrt{1-\frac{4m_{e}^{2}}{m_{\varphi}^{2}}} \label{phiee}%
\end{align}

From this clean expression, assuming $m_{\varphi}\geq10$ MeV, we estimate
\begin{equation}
\delta_{e}\leq\sqrt{4\pi H_{0}/m_{\varphi}}\sim10^{-20}%
\end{equation}
which implies the constraint
\begin{equation}
Y\leq Y_{\max}\text{, }Y_{\max}\sim10^{26}m_{\varphi}^{-5/2}%
\end{equation}
This becomes meaningful only if $m_{\varphi}\geq10$ GeV, since $Y_{\max
}(m_{\varphi}=10$ GeV$)\approx5$.

\bigskip

$\mathbf{\varphi\rightarrow2\gamma}$. Quintessence can always decay
radiatively into two photons, analogous to the Higgs decay into two photons.
The Higgs decay rate is well known \cite{Higgs decay},%
\begin{equation}
\Gamma(h\rightarrow2\gamma)=\frac{G\alpha^{2}M_{h}^{3}}{128\sqrt{2}\pi^{3}%
}\left\vert A_{1}(\tau_{W})+\sum N_{c}Q_{f}^{2}A_{1/2}(\tau_{f})\right\vert
^{2}%
\end{equation}
where $G$ is the Fermi constant, $\alpha$ the fine-structure constant, $N_{c}$
the color factor, $Q_{f}$ the fermion charge, and , $\tau_{i}=M_{h}^{2}%
/4M_{i}^{2}$ with $i=f,W$. The loop functions are:%
\begin{align}
A_{1}(\tau)  &  =-\left(  2+3/\tau+3(2/\tau-1/\tau^{2})f(\tau)\right) \\
A_{1/2}(\tau_{f})  &  =2\left(  \tau+(\tau-1)f(\tau)\right)  /\tau^{2}%
\end{align}
with
\begin{align}
\tau &  \leq1\text{, \ \ \ }f(\tau)=\arcsin^{2}\sqrt{\tau}\\
\tau &  >1\text{, \ \ \ }f(\tau)=-\frac{1}{4}\left[  \ln\frac{\sqrt{\tau
}+\sqrt{\tau-1}}{\sqrt{\tau}-\sqrt{\tau-1}}-i\pi\right]  ^{2}%
\end{align}

To obtain $\Gamma_{\varphi\rightarrow2\gamma}$, quintessence decay rate, we
should replace $M_{h}$ with $m_{\varphi}$ where is a need and adjust
couplings. We note that typically $m_{\varphi}$ is light compared to heavy
fermions and W bosons, so $\tau_{W}\ll1$ and $\tau_{f}\ll1$. This gives the
approximations%
\begin{equation}
A_{1}(\tau_{W})\approx-7\text{, \ \ \ \ \ }A_{1/2}(\tau_{f})\approx4/3
\end{equation}
For very light quintessence, all fermions contribute similarly. If four
momentum of the quintessence is $k_{\mu}$%
\begin{align}
\Gamma_{\varphi\rightarrow2\gamma}  &  =\frac{\bar{\delta}^{2}G\alpha
^{2}(k_{\mu}^{2})^{2}}{128\sqrt{2}\pi^{3}k_{0}}\left\vert -7+\frac{4}{3}\sum
N_{c}Q_{f}^{2}\right\vert ^{2}\\
\bar{\delta}  &  =\delta\frac{m_{\varphi}^{2}}{M_{h}^{2}-m_{\varphi}^{2}}%
\end{align}

Since $\sum N_{c}Q_{f}^{2}=8$, for the rest frame quintessence decay rate we
should have
\begin{equation}
\Gamma_{\varphi\rightarrow2\gamma}=\frac{\bar{\delta}^{2}G\alpha^{2}%
m_{\varphi}^{3}}{128\sqrt{2}\pi^{3}}\left[  11/3\right]  ^{2}\sim10^{-30}%
\bar{\delta}^{2}m_{\varphi}\left(  \frac{m_{\varphi}}{1eV}\right)  ^{2}%
\end{equation}
Comparing this to the Hubble constant, we find
\begin{equation}
\bar{\delta}^{2}\left(  \frac{m_{\varphi}}{1\text{ eV}}\right)  ^{3}%
\sim10^{-3}\rightarrow\delta^{2}\frac{m_{\varphi}^{4}}{M_{h}^{4}}\left(
\frac{m_{\varphi}}{1\text{ eV}}\right)  ^{3}\sim10^{-3}%
\end{equation}

Since $\delta$ is at leasy $10^{-14}$, being proportional to the ratio of the
active neutrino mass to the electroweak scale, this constraint may become
cosmologically relevant if $m_{\varphi}>1MeV$. For $m_{\varphi}>1MeV$, lighter
fermions no longer contribute significantly, but the overall magnitude of the
decay rate remains unchanged.

Here we compared the tiny decay rate of quintessence to $H_{0}$. However, if
values below $H_{0}$ are preferred \cite{chanda}, the $\ $($Y$, $m_{\varphi}$)
parameter space becomes more narrow.

\section{The mass mixing}

When the mixing term is introduced as
\begin{equation}
L_{mix}=4\delta M^{2}h\varphi
\end{equation}
an orthogonal transformation linear in $\delta$%
\begin{equation}
h\rightarrow h-4\delta\frac{M^{2}}{M_{h}^{2}-m_{\varphi}^{2}}\varphi;\text{
\ }\varphi\rightarrow\varphi+4\delta\frac{M^{2}}{M_{h}^{2}-m_{\varphi}^{2}}h
\label{MMT}%
\end{equation}
diagonalizes the scalar states. Here the transformation parameter depends
explicitly on the sterile neutrino mass $M$. Since in this case the sterile
neutrino is comparatively light, the mixing is additionally suppressed. The
contribution to the diagonal mass values is again a second order effect,
proportional to $\delta^{2}$. \ New interactions appear in analogy to the
kinetic mixing case, but are further suppressed by the ratio of the sterile
neutrino mass to the heavier of the Higgs and quintessence masses.

Unlike the kinetic mixing scenario, processes requiring light quintessence
(such as weak and Higgs boson decays considered above) do not occur here.
However, quintessence decay processes are amplified by the large quintessence mass.

\textbf{Quintessence decay into }$\mathbf{e}^{+}\mathbf{e}^{-}$. The decay
rate into an electron--positron pair is modified directly from (\ref{phiee})
by replacing
\begin{equation}
\delta_{e}\rightarrow4\delta\dfrac{\mathrm{y}_{e}M^{2}}{M_{h}^{2}-m_{\varphi
}^{2}}\text{ }%
\end{equation}
This modifies the restriction to
\[
\delta\leq\frac{\left\vert M_{h}^{2}-m_{\varphi}^{2}\right\vert \sqrt{\pi
H_{0}/4m_{\varphi}}}{\mathrm{y}_{e}M^{2}}%
\]
numerically yelding
\begin{align}
\delta(m_{\varphi}  &  =0.1\text{ TeV},M=1\text{ GeV})\leq10^{-13}\\
\delta(m_{\varphi}  &  =1\text{ TeV},M=1\text{ GeV})\leq10^{-11}%
\end{align}
This sets a constraint on the Yukawa coupling $Y,$ which is of order unity for
$m_{\varphi}=0.1$ TeV, $M=1$ GeV. This is essentially equivalent to the limit
derived in the previous subsection. For heavier quintessence, the constraint
becomes considerably relaxed. In any case, this is not a primary decay channel
for quintessence. Compared to the dominant sterile neutrino decay rate
(\ref{phiNN}), it is completely negligible -- unlike the light quintessence
scenarios, where the sterile neutrino decay channel is kinematically forbidden.%

\begin{equation}
\frac{\Gamma_{\varphi\rightarrow ee}}{\Gamma_{\varphi\rightarrow NN}}%
\sim\dfrac{\delta^{2}\mathrm{y}_{e}^{2}M^{4}}{Y^{2}(M_{h}^{2}-m_{\varphi}%
^{2})^{2}}\leq\frac{m_{\nu}^{2}m_{e}^{2}M^{4}}{V_{EW}^{4}(M_{h}^{2}%
-m_{\varphi}^{2})^{2}}%
\end{equation}

\textbf{Quintessence decay into photons.} For the decay channel into photons,
when quintessence is assumed to be much heavier than the Higgs, the loop
contributions simplify:
\[
A_{1}(\tau_{W})\approx-2,A_{1/2}(\tau_{f})\approx0
\]
The decay rate is then estimated as%
\begin{equation}
\Gamma(h\rightarrow2\gamma)\sim\delta^{2}\frac{M^{4}}{m_{\varphi}}%
\frac{G\alpha^{2}}{\pi^{3}}\sim\delta^{2}10^{-5}\left(  \frac{M}{1\text{GeV}%
}\right)  ^{4}\left(  \frac{1\text{TeV}}{m_{\varphi}}\right)  \text{ }\left(
\text{eV}\right)
\end{equation}
This decay rate is also strongly suppressed compared to (\ref{phiNN}) and
appears unimportant.

\section{Momentum-dependent mixing}

When the parameter space or energy scale of the process does not allow a
blanket diagonalization of the scalar fields, the off diagonal self-energy
depends nontrivially on momentum. In such cases, a single momentum independent
diagonalization of the fields is not possible, and one must instead account
for the effects of $\Pi_{h\varphi}(p_{\mu}^{2})$ directly at the Feynman
diagram level.

For example, consider an on-shell Higgs decaying through oscillation into
quintessence, followed by quintessence decaying into some final state $X$. The
matrix element for the decay rate in the rest frame is
\begin{equation}
\mathcal{M}_{h\rightarrow X}=\Pi_{h\varphi}(k_{\mu}^{2})\frac{i}{k_{\mu}%
^{2}-m_{\varphi}^{2}}\mathcal{M}_{\varphi\rightarrow X}=\frac{i\Pi_{h\varphi
}(M_{h}^{2})}{M_{h}^{2}-m_{\varphi}^{2}}\mathcal{M}_{\varphi\rightarrow X}%
\end{equation}

\begin{itemize}
\item Light sterile neutrino case ($M_{h}^{2}\gg M^{2}$): $\Pi_{h\varphi
}(M_{h}^{2})\approx4i\delta M^{2}$ (\ref{MM}), giving:%
\begin{equation}
\mathcal{M}_{h\rightarrow X}\approx\frac{-4\delta M^{2}}{M_{h}^{2}-m_{\varphi
}^{2}}\mathcal{M}_{\varphi\rightarrow X}%
\end{equation}
which reproduces the mass mixing behavior.

\item Heavy sterile neutrino case ($M_{h}^{2}\ll M^{2})$: $\Pi_{h\varphi
}(M_{h}^{2})\approx i\delta M_{h}^{2}$, giving
\begin{equation}
\mathcal{M}_{h\rightarrow X}\approx\frac{-\delta M_{h}^{2}}{M_{h}%
^{2}-m_{\varphi}^{2}}\mathcal{M}_{\varphi\rightarrow X}%
\end{equation}
which mimics the kinetic mixing behavior (\ref{km1}).
\end{itemize}

When the situation is reversed --- quintessence decaying through oscillation
into the Higgs, followed by Higgs decay into some final state $\overline{X}$
-- the amplitude is
\begin{equation}
\mathcal{M}_{\varphi\rightarrow\overline{X}}=\Pi_{h\varphi}(k_{\mu}^{2}%
)\frac{i}{k_{\mu}^{2}-M_{h}^{2}}\mathcal{M}_{h\rightarrow\overline{X}}%
=\frac{i\Pi_{h\varphi}(m_{\varphi}^{2})}{m_{\varphi}^{2}-M_{h}^{2}}%
\mathcal{M}_{h\rightarrow\overline{X}}%
\end{equation}
again reproducing either kinetic or mass mixing behavior depending on the
hierarchy of scales.

Because of possible hierarchies between mass scales, we can encounter mixed
behavior. When there is no clear hierarchy, or when the propagating momentum
is off-shell, the full expression must be retained and evaluated case-by-case.

This non-uniform regime does not introduce fundamentally new phenomena; it
simply interpolates between the properties of mass mixing and kinetic mixing
depending on the process. The approach described here is universally valid but
requires case-by-case consideration, whereas the pure mass mixing or kinetic
mixing limits offer more simplicity and uniformity when the parameter space
allows them.

\bigskip

\section{Conclusion}

We investigated the consequences of coupling a quintessence field to sterile
neutrinos in a seesaw framework. While the interaction between quintessence
and sterile neutrinos is introduced at tree level, the presence of the
standard Higgs--sterile-neutrino Yukawa interaction inevitably generates Higgs
-- quintessence mixing through radiative corrections. This effect provides a
portal through which quintessence acquires effective interactions with
Standard Model particles.

The one-loop mixing amplitude was calculated and analysed in different
momentum regimes. We find that the nature of the mixing depends on the
relation between the characteristic momentum transfer and the sterile-neutrino
mass. For $p%
%TCIMACRO{\U{b2}}%
%BeginExpansion
{{}^2}%
%EndExpansion
\ll M%
%TCIMACRO{\U{b2}}%
%BeginExpansion
{{}^2}%
%EndExpansion
$ the mixing assumes a kinetic form and can be represented by an effective
derivative operator between the Higgs and quintessence fields. In contrast,
for $p%
%TCIMACRO{\U{b2}}%
%BeginExpansion
{{}^2}%
%EndExpansion
\gg M%
%TCIMACRO{\U{b2}}%
%BeginExpansion
{{}^2}%
%EndExpansion
$ the mixing behaves as a mass-type contribution. Between these limits the
mixing remains momentum dependent and cannot generally be removed by a single field redefinition.
The induced Higgs--quintessence mixing generates effective couplings of
quintessence to fermions and gauge bosons. As a result, processes involving Higgs, $W$ and $Z$ bosons receive additional contributions through intermediate quintessence states. The corresponding corrections are strongly suppressed by neutrino masses and by loop factors and are therefore expected to remain below current experimental sensitivities in the minimal scenario considered here.

The same mixing mechanism opens decay channels for quintessence. For
sufficiently heavy quintessence fields, decays into charged fermions become
possible, while radiative decay into two photons remains available even for
very light quintessence. Although these processes are highly suppressed, the
extreme longevity required of a dark-energy field makes even tiny decay rates
potentially relevant from a cosmological perspective. The resulting
constraints provide a complementary way of probing interactions between
quintessence and sterile neutrinos.

The suppression obtained in the present work is largely a consequence of the
heavy-seesaw limit. In models containing several sterile neutrino states,
lighter sterile neutrinos may contribute significantly to loop processes and
modify the phenomenology. In such cases the momentum dependence of the mixing
and threshold effects can become considerably more important, potentially
enhancing observable signatures.

Overall, the results demonstrate that sterile-neutrino-assisted quintessence
models naturally predict Higgs--quintessence mixing and consequently a broad
class of effective interactions between quintessence and Standard Model
fields. While the effects appear small in the simplest heavy-seesaw
realization, the framework establishes a direct connection between dark-energy
dynamics, neutrino mass generation and collider-scale particle physics.

\bigskip

%

%TCIMACRO{\TeXButton{TeX field}{\section*{Acknowledgments}}}%
%BeginExpansion
\section*{Acknowledgments}%
%EndExpansion

I would like to thank Jon Chkareuli and Michael Maziashvili for useful and
fruitful discussions.

\end{document}